\documentclass[12pt]{article}

\usepackage[english]{babel}

\usepackage[letterpaper,top=2cm,bottom=2cm,left=3cm,right=3cm,marginparwidth=1.75cm]{geometry}

\usepackage{amsmath}
\usepackage{graphicx}
\usepackage[colorlinks=true, allcolors=blue]{hyperref}
\usepackage{amssymb}
\usepackage[utf8]{inputenc}
\usepackage{lipsum}
\usepackage{multicol}
\usepackage{physics}

\title{A novel pre-inflationary model in view of the lack of angular correlation of CMB}
\author{ M. Montes $^{1}$, Jos\'e Edgar Madriz Aguilar$^{1}$$^{\dagger}$, A. Bernal$^{1}$,   Diego Allan Reyna $^{2}$, }

\begin{document}
	\date{}
	\maketitle
	\begin{center}
		$^{1}$ Departamento de Matem\'{a}ticas, Centro Universitario de Ciencias Exactas e ingenier\'{i}as (CUCEI),
		Universidad de Guadalajara (UdG), Av. Revoluci\'on 1500 S.R. 44430, Guadalajara, Jalisco, M\'exico,\\
		and\\
		$^{2}$ Department of Physics, McGill University, 3600 Rue University, Montréal, H3A 2T8, QC Canada \\
		\vspace{0.3cm}
		
		$^{\dagger}$ Corresponding author: José Edgar Madriz Aguilar\\

		E-mails:  jose.madriz@academicos.udg.mx, 
		alfonso.bernal@academicos.udg.mx,
		allan.reyna@mail.mcgill.ca, mariana.montes@academicos.udg.mx
	\end{center}
	\begin{abstract}
        In this paper we propose a novel unified cosmological model that connects a pre-inflationary epoch, starting at the Planckian time, with the onset of inflation within a single scalar-field framework. The pre-inflationary phase is characterized by a decelerated expansion with an increasing comoving Hubble horizon, followed by a gradually transition to an accelerated inflationary regime. This early dy\-na\-mics leads to a modified causal structure that naturally accounts for the suppression of large-angle $(\theta \gtrsim 60^\circ)$ correlations in the cosmic microwave background (CMB) reported by the satellite PLANCK. We study the quantum fluctuations of the scalar field using the Mukhanov--Sasaki formalism and a canonical quantization procedure based on energy minimization. We find that the vacuum state is well-defined only for sub-horizon modes at the onset of inflation, which induces a natural cutoff in the primordial power spectrum. The resulting spectrum exhibits a suppression at large scales while remaining nearly scale-invariant at small scales. In the appropriate limit, the model recovers the standard de Sitter result, in agreement with current observational constraints. These results highlight the relevance of pre-inflationary dynamics for addressing large-scale anomalies within a consistent inflationary framework.
	\end{abstract}
	
	\\
	\vspace{0.3cm}
	
	
	\section{\large{Introduction}}
	
	The cosmic microwave background radiation (CMB) has been a cornerstone of our current understanding of the universe. Observations of the CMB, made particularly by space-based missions such as COBE, WMAP, and Planck\cite{Sm99, Hin13, PlanckVII}, have provided a wealth of information about its earlier evolution and its large-scale properties revealing that the universe is, on large scales, homogeneous, isotropic, and spatially flat. Moreover, one of the most significant achievements made by these missions is the high-precision mapping of the anisotropies present in the CMB. To account for the aforementioned large-scale characteristics of the universe, cosmological inflationary models were introduced in the late 20th century \cite{Gu81, Lin82}. The standard inflationary paradigm postulates the emergence of an inflaton field at the Planck time $t_p$, which generates su\-ffi\-cient vacuum energy to drive a period of accelerated expansion in the early universe \cite{Lin82}. This inflationary epoch explains the observed homogeneity, isotropy, and flatness of the universe \cite{Lin83}. Additionally, through the quantum vacuum fluctuations of the inflaton field, inflation offers a mechanism capable of explaining the origin of the primordial anisotropies observed in the CMB \cite{MuCh81, Muk}.\\
	
	The Planck 2018 results are highly consistent with slow roll inflation. Furthermore, these observations provide compelling evidence for inflationary scenarios with plateau-like potentials, which naturally reconcile the observed spectral tilt of primordial fluctuations with the tight constraints on tensor modes \cite{PlanckX, IjStLo13}. However, despite the success of the scalar field slow-roll inflation framework, it still faces several cha\-llen\-ges. As previously mentioned, Planck data seems to favor plateau-like potentials. However, it has been shown that a slow-roll inflaton with a plateau-like potential is incompatible with the emergence of an inflaton field characterized by a Planck-scale energy density $\rho_p$ at the Planck time $t_p$ \cite{LiuMelia2024}. More precisely, it has been found that the energy density associated with the plateau of the potential is 12 orders of magnitude below the Planck scale \cite{IjStLo13}.\\
	
	Even more, the Planck mission has observed a lack of correlation in the two-point angular correlation function at large angular separations ($\theta \gtrsim 60^\circ$)\cite{PlanckVII}. This feature has been related to the emergence of a cutoff in the primordial power spectrum and to a delayed onset of inflation \cite{LiuMelia2024, MeLo18}. This delay of inflation to begin, raises the possibility of an intermediate phase existing between the Planck epoch and the inflationary period, sometimes called the pre-inflationary epoch. In addition to this, a suppression of power in the curvature fluctuations at large scales has been observed \cite{Ag17}. To address these issues, the existence of a kinetically dominated phase during pre-inflation has been proposed \cite{CoPeKoLi03, Handley2014, MeAl20}.\\
	
	The observational picture described previously poses a problem for the quantum initial conditions of the inflaton field fluctuations. These fluctuations are usually assumed to be in a Bunch–Davies vacuum state; however, this choice  is only valid in a FLRW space-time in which the modes obey the condition  $k/aH\ll1$ in the UV sector, as happens for example in a De-Sitter inflation. In a general FLRW background—such as the one describing a kinetically dominated universe, this assumption is no longer applicable, moreover, the notion of a vacuum state becomes ambiguous \cite{ArLi03}. In an effort to resolve these limitations, alternative vacuum selection prescriptions have been explored,  as for instead Hamiltonian diagonalization, adiabatic vacua, and the minimization of the renormalized stress-energy tensor \cite{ArLi03, HaLaHo16}.\\

	In this work, we propose a new cosmological model of the early universe in which a pre-inflationary period, starting at planckian time, gradually evolves from a decelerated to an accelerated expansion until it reaches a quasi-de Sitter inflationary period. We identify the  presence of the previously mentioned  cutoff in the power spectrum  with a transition epoch from a decelerated expansion (pre-inflation) to an accelerated one (inflation). The paper is organized as follows. In section 1 we give a brief introduction. In section 2 we introduce the Mukhanov-Sasaki equation and the canonical quantization procedure employed in the standard inflationary formalism. Section 3 is devoted to the normalization procedure via minimization of the energy density expectation value. Section 4 examines the behavior of the quantum fluctuations  related to the existence of a cut-off in a period of decelerated expansion. In section 5, we show the existence of a cut-off in the power spectrum in a delayed inflation stage.  Section 6 is dedicated to the construction of our unified model of pre-inflation and delayed inflation. In section 7 we obtain the dynamics of the quantum fluctuations of the scalar field during pre-inflation and the delayed inflation stages. Finally in section 8 we give some final comments.
    In our conventions we will work in natural units $c = \hbar = 1 $.

	\section{Standard Inflation}
    
Before explaining our model of the early universe that describes in an unified manner the pre-inflationary and the inflationary epochs, let us firstly, in this section, focus on some important concepts involved in the development of the model. The model will be addressed in section 3. \\

With the aim of investigating the effects of the expansion background dynamics on the vacuum choice, and its relation to the primordial inflaton quantum fluctuations power spectrum, in this section we will establish the relevant equations and the outline of the standard inflationary formalism. In the conventional inflationary framework, at Planck time $t_p$ the Universe is assumed to be dominated by a scalar field minimally coupled to gravity. Hence, the action corresponding to standard inflation is \begin{equation}\label{ac_1}
		S = \int d^4 \sqrt{-g} \left[ \frac{R}{16 \pi G} - \frac{1}{2} g^{\mu\nu} \partial_{\mu}\phi\partial_{\nu}\phi + U(\phi)\right],
	\end{equation}
	where  $g_{\mu\nu}$ is the metric tensor, $R$ denotes the Ricci scalar, $\phi$ the inflaton field and $U(\phi)$ its potential. Applying Hamilton´s principle, it is straightforward to show that the field equations derived from the action \eqref{ac_1} correspond to the Einstein equations and the Klein-Gordon equation
	
	\begin{equation}\label{Einstein_1}
		R_{\mu \nu} - \frac{1}{2} g_{\mu\nu} R = 8\pi G \left[\phi,_{\mu} \phi,_{\nu} - \frac{g_{\mu \nu}}{2}\left(\phi_{,\alpha}\phi^{,\alpha} - 2 U(\phi)\right)\right],
	\end{equation}
	\begin{equation}\label{klein_1}
		\Box \phi + \frac{dU}{d\phi} = 0,
	\end{equation}
with $R_{\mu \nu}$ denoting the Ricci curvature tensor and $\Box=g^{\mu\nu}\nabla_{\mu}\nabla_{\nu}$ representing the D'Alambertian , and the coma denotes a partial derivative.\\

At this point, due to the cosmological principle, an spatially homogeneous and isotropic universe is assumed. In order to respect this principle, the metric we consider is the one in the spatially flat Friedmann-Lemaitre-Robertson-Walker (FLRW) line element that reads
	\begin{equation}\label{FLRW}
		ds^2 = dt^2 - a^2(t)(dx^2 + dy^2 + dz^2),
	\end{equation}
	where $a(t)$ is the cosmic scale factor, $t$ is the cosmic time and $\lbrace x,y,z\rbrace$ are the cartesian coordinates. From the equations \eqref{Einstein_1}, \eqref{klein_1} and \eqref{FLRW} we arrive to the expressions
	\begin{equation}\label{friedmann_1}
		3H^2 = 8\pi G \rho,
	\end{equation}
	\begin{equation}\label{klein_2}
		\ddot{\phi} + 3H \dot{\phi}-\frac{1}{a^2}\nabla^2\phi + \frac{dU(\phi)}{d\phi} = 0,
	\end{equation}
	with $H = \dot{a}/a$ being the Hubble parameter, and $\rho$ is the energy density of the inflaton field that in our case has the form
	\begin{equation}\label{energy_density_1}
		\rho=\frac{1}{2}\dot{\phi}^2-\frac{1}{2a^2}(\nabla\phi)^2+U(\phi)\,,
	\end{equation}
	where the dot represents derivative with respect to the time $t$. \\

	In order to describe the quantum vacuum fluctuations of the inflaton field related to the primordial anisotropies observed in the CMB, we will consider scalar fluctuations of the metric in the spatially flat gauge, for which the line element reads
	\begin{equation}\label{semiclassical_ds}
		ds^2 = a^2(\tau) \left[ (1 + 2A)\, d\tau^2 - 2B_{,i}\, dx^i d\tau - \delta_{ij}\, dx^i dx^j \right],
	\end{equation}
where $A = A(\tau, \bar{x})$ and $B=B(\tau, \bar{x})$ represent the scalar fluctuations of the metric; and $\tau$ is the conformal time  introduced to simplify the analysis and clearer physical interpretation of perturbation dynamics. Conformal time is widely used in inflationary perturbation theory because it makes the expanding spacetime conformally equivalent to Minkowski space. This simplifies the structure of wave equations for perturbations, turning them into flat-space–like equations with a time-dependent effective mass. In particular, the Mukhanov–Sasaki equation takes the form of a canonical harmonic oscillator, which is crucial for quantization. Conformal time also aligns causal structure with light-cone propagation, making horizon crossing  more transparent. Additionally, mode functions behave as simple plane waves in the sub-horizon limit, facilitating the definition of the Bunch–Davies vacuum. In our case, the conformal time is given by
	\begin{equation}\label{conformal_time}
		\tau = \int \frac{dt}{a(t)}.
	\end{equation}
	Similarly, due to the semiclassical approximation we have adopted, the inflaton field can be expressed in the form
	\begin{equation}\label{semiclassical_phi}
		\phi (\tau, \bar{x}) = \phi_b (\tau)  + \frac{f(\tau, \bar{x})}{a(\tau)},
	\end{equation}
	where $f(\tau, \bar{x})=$ describes the comoving perturbations of the inflaton field, $\phi_b$ = $\ev{\phi}$  is the classical part of the inflaton field $\phi$ that obeys the cosmological principle, valid only on cosmological large scales. Besides, the quantum fluctuations of the inflaton field $\delta\phi=f(\tau,\bar{x})/a(\tau)$  must obey the conditions: 
     $\expval{\delta\phi}=\expval{\delta\phi^{'}}=0$ and $\left<\delta\phi^2\right>\neq 0$, with $\expval{\,}$ denoting the expectation value with respect to some physical quantum state $\ket{E}$ for $\phi$, and the prime denoting derivative with respect to the conformal time $\tau$.
     \\
     
By applying the perturbed metric \eqref{semiclassical_ds} in the equation \eqref{klein_1}, we find that the dynamical equation for the quantum fluctuations of the inflaton field reads
     \begin{equation}\label{qcg1}
         \delta\phi^{\prime\prime}+2{\cal H}\delta\phi^{\prime}-\nabla^{2}\delta\phi=(A^{\prime}+\nabla^2 B)\phi_{b}^{\prime}-2a^2 U_{,\phi}A-a^2 U_{,\phi\phi}\delta\phi\,,
     \end{equation}
    where ${\cal H}=a^{\prime}(\tau)/a(\tau)$ is the comoving Hubble parameter and $U_{,\phi\phi}=\frac{d^2U}{d\phi^2}$. Now, it follows form the perturbed Einstein equations $\delta G_{00}=8\pi G \delta T_{00}$ and $\delta G_{0i}=8\pi G \delta T_{0i}$ the relations
    \begin{eqnarray}\label{qcg2}
        A &=& 4\pi G \frac{\phi_b^{\prime}}{{\cal H}}\delta\phi\,,\\
        \label{qcg3}
        \nabla^2 B &=& -4\pi G\frac{\phi_b^{\prime}}{{\cal H}}\left[\delta\phi^{\prime}+\left(1-\frac{\phi_b^{\prime\prime}}{{\cal H}\phi_b^{\prime}}-4\pi G\frac{(\phi_{b}^{\prime})^2}{{\cal H}^2}\right){\cal H}\delta\phi\right]\,.
    \end{eqnarray}
    Now, inserting \eqref{qcg2} and \eqref{qcg3} in \eqref{qcg1} and with the help of \eqref{semiclassical_phi},  we obtain
	\begin{equation}\label{mukhanov_1}
		f^{''} - \nabla^2 f - \frac{z^{''}}{z}f = 0,
	\end{equation}
	where $z \equiv a\phi_{b}^{'}/\mathcal{H}$. This equation is known as the Mukhanov-Sasaki equation and it determines the dynamics of the comoving fluctuations of the inflaton field coupled to the scalar perturbations.\\

    Once we have revised the equation that determines the dynamics of the comoving fluctuations of the inflaton coupled with the scalar perturbations of the metric, in the spatially flat gauge, we will now focus in reviewing the canonical quantization procedure commonly applied to these perturbations. \\

    At the beginning of inflation, the relevant modes are deep inside the Hubble horizon, and as inflation proceeds, they are stretched beyond it. Modes inside the horizon are named sub-horizon modes, or ultraviolet UV sector. When the modes are outside the horizon, they freeze, and this is called the infrared (IR) sector.  In the UV sector the dynamics of the sub-horizon modes is approximately that of quantum fields in flat spacetime, allowing one to define vacuum states through the behavior of harmonic oscillators and to directly apply the uncertainty principle. \\

    In this manner, let us to begin by imposing the uncertainty principle by means of the commutation relation
	\begin{equation}\label{commutation_I}
		\left[\hat{f}(\tau, \bar{x}), \hat{\Pi}^{(f)}(\tau, \bar{x'})\right] = i \delta(\bar{x} - \bar{x'}),
	\end{equation} 
	where $\hat{\Pi}^{(f)} \equiv \partial \mathcal{L}/\partial \hat{f}'$ is the canonical conjugate momentum to $f$.
Following the standard procedure, we express the comoving fluctuations $f(\tau, \bar{x})$ as a continuous sum of Fourier modes $ f_k(\tau)$  in the form
	\begin{equation}\label{fourier_f}
		f(\tau, \bar{x}) = \int  \frac{d^3k}{(2\pi)^\frac{3}{2}}\left[\hat{a}_k  f_k(\tau) e^{i\bar{k}\cdot\bar{x}} + \hat{a}^{\dagger}_k  f_k^{*}(\tau) e^{-i\bar{k}\cdot\bar{x}} \right],
	\end{equation}
	where $\hat{a}_k$ and $\hat{a}^{\dagger}_k$ denote the annihilation and creation operators, respectively, and are required to satisfy the commutation algebra
	\begin{equation}\label{ca1}
		\left[\hat{a}_k, \hat{a}^{\dagger}_{k'} \right] = \delta(\bar k - \bar k'),
	\end{equation}
	\begin{equation}\label{ca2}
		\left[\hat{a}_k, \hat{a}_{k'} \right] = [\hat{a}^{\dagger}_k, \hat{a}^{\dagger}_{k'}] = 0.
	\end{equation}
    It is important to mention that here $\dagger$ means transposed complex conjugate whereas the asterisk accounts for complex conjugate. We highlight that the prime in the relations \eqref{ca1} and \eqref{ca2} is used only to express that $\bar{k}$ and $\bar{k}^{\prime}$ are two different momentum vectors, and has nothing to do with derivatives with respect to the conformal time. \\
    
Now, inserting \eqref{fourier_f} in \eqref{mukhanov_1}, it is not difficult to show that the modes $f_k$ satisfy the equation
	\begin{equation}\label{mukhanov_2}
		f^{''}_k + \left(k^2  - \frac{z^{''}}{z}\right)f = 0.
	\end{equation} 
On the other hand, by applying the semiclassical approximation \eqref{semiclassical_ds} and \eqref{semiclassical_phi} to the action \eqref{ac_1}, and retaining terms up to second order, it is straightforward to obtain that the action for the fluctuations $f(\tau, \bar{x})$ in the flat gauge takes the form
	\begin{equation}\label{ac_f}
		S_{(f)} = -  \frac{1}{2} \int d\tau d^3x \, \left[(f^{'})^2 - \left(\nabla f\right)^2 - \frac{z^{''}}{z} f^2 \right].
	\end{equation}
	
	Taking into account the action \eqref{ac_f} and the definition of $\hat{\Pi}^{(f)}$, the commutation relation \eqref{commutation_I} reduces to 
	\begin{equation}\label{commutation_II}
		\left[\hat{f}(\tau, \bar{x}), \hat{f'}(\tau, \bar{x'})\right] = i \delta(\bar{x} - \bar{x'}).
	\end{equation} 
	Thus, employing the Fourier expansion \eqref{fourier_f} in the relation \eqref{commutation_II}, one finds that the mode functions $f_k(\tau)$ must satisfy the normalization condition
	\begin{equation}\label{normalization_cond}
		f_k f_k^{'*} - f^{*}_k f'_k = i.
	\end{equation}
	The normalization condition \eqref{normalization_cond} is not sufficient to fully specify the solutions $f_k$ of the second order Mukhanov equation \eqref{mukhanov_2}.  To this end, it will be necessary to choose a vacuum state $\ket{0}$. This leaves us with the question of how to make an appropriate choice. In the next section, we will have a little discussion about this issue.
	
	\section{Normalization via minimization of 
\texorpdfstring{$\expval{\hat{T}_{00}}{0}$}{Expectation value of T00 in vacuum}}

    The vacuum state $\ket{0}$, defined as $\hat{a}_k\ket{0} = 0,$ for all $\bar{k}$, plays a central role in canonical quantization. In quantum field theory in flat space-time, the notion of particles is intrinsically linked to the vacuum, which is typically identified as the state of lowest energy. In curved space-time, however, there is no unique prescription for selecting a preferred vacuum state, although it remains a fundamental element of the formalism. This ambiguity becomes particularly relevant in inflationary cosmology, where the predicted power spectrum of primordial fluctuations depends sensitively on the choice of vacuum. To analyze this dependence, it is common to approximate the inflationary phase by a de Sitter background, which captures the essential features of the formalism.\\
	
Using the conformal time, the scale factor in a de Sitter expansion is expressed as
	\begin{equation}\label{de_Sitter}
		a(\tau) = - \frac{1}{H \tau}.
	\end{equation}
	Here, it is important to remember that  $-\infty < \tau < 0$. Considering the scale factor \eqref{de_Sitter}, the Mukhanov equation \eqref{mukhanov_2}, in a De Sitter epoch, takes the form
	\begin{equation}\label{mukhanov_3}
		f^{''}_k + \left(k^2  - \frac{2}{\tau^2}\right)f = 0.
	\end{equation} 
The general solution for \eqref{mukhanov_3} reads
	\begin{equation}\label{sol_1}
		f_k(\tau) =  \frac{A_k}{\sqrt{2k}} \left(1 -\frac{i}{k \tau}\right) e^{-ik\tau} +  \frac{B_k}{\sqrt{2k}} \left(1 + \frac{i}{k \tau}\right) e^{ik\tau}\,,
	\end{equation}
	where $A_k$ and $B_k$ are free constants, which are determined by the normalization condition \eqref{normalization_cond} and the selection of a vacuum state. Substituting the solution \eqref{sol_1} in the normalization condition \eqref{normalization_cond}, it follows that 
	\begin{equation}
		|A_{k}|^2 - |B_k|^2 = 1.
	\end{equation}
	
	As widely established in the literature, the power spectrum $P_{\delta \phi}(k)$ of the fluctuations $\delta \phi(\tau, \bar{x}) = f(\tau, \bar{x})/a(\tau)$ is related to their two-point expectation value of the field $f$ as follows
	\begin{equation}\label{power_I}
		\bra{0} \hat{f}^2(\tau, \bar{x})\ket{0} = \int_{k_0}^{k_{max}} \, d\left(\ln k\right) \frac{k^3}{2\pi^2} |f_k(\tau)|^2 = \int_{k_0}^{k_{max}} d\left(\ln k\right)\, a^2(\tau) P_{\delta \phi}(k)\,,
	\end{equation}
    where $k_0$ denotes an infrared cutoff, representing the minimum comoving wavenumber (largest physical scale) included in the spectrum. In inflationary cosmology, the quantity of interest is the power spectrum at the end of inflation, corresponding to the IR limit, which can be expressed as 
	\begin{equation}\label{power_II}
		\left. P_{\delta \phi}(k) \right\rvert_{IR} = \lim_{\tau \to 0}  \frac{k^3}{2\pi^2} \frac{|f_k(\tau)|^2}{a^2(\tau)}.
	\end{equation}	
This expression represents the power spectrum of scalar field fluctuations evaluated in the infrared (super-horizon) limit. Physically, the power spectrum $P_{\delta \phi}(k)$ quantifies the amplitude of quantum fluctuations of the inflaton field as a function of the comoving wavenumber $k$. In this regime, modes have exited the Hubble radius and their amplitudes become effectively frozen, providing the seeds for the later formation of large-scale structure in the universe.\\

Now, substituting the solution \eqref{sol_1} into the power spectrum in the IR-sector \eqref{power_II}, we obtain that at the end of inflation, the power spectrum associated with the fluctuations of the inflaton field $(\delta \phi = f/a)$ has the form
	\begin{equation}\label{ps_end}
		\left. P_{\delta \phi}(k) \right\rvert_{IR} = \frac{H^2}{4\pi^2}|A_k - B_k|^2.
	\end{equation}
It follows from the equation \eqref{ps_end} that the power spectrum strongly depends on the choice of the constants $A_k$ and $B_k$, and therefore on the choice of  vacuum. Since an inflationary epoch corresponds to a quasi-De Sitter space-time, a natural vacuum choice is the Bunch-Davies vacuum \cite{BDp}. This vacuum state is invariant under the De Sitter symmetry group, and corresponds to the selection $A_k=1$ $B_k=0$.\\

Due to recent observational evidence suggesting a delayed onset of inflation, it is possible that the beginning of inflation does not correspond to a de Sitter spacetime. Consequently, the assumption of a Bunch–Davies vacuum as the initial condition for inflaton fluctuations may no longer be applicable.
In order to solve this problem, different vacuum choices have been proposed. Particular attention has been given to the standard Hamiltonian diagonalization approach. This procedure selects the vacuum state that minimizes the expectation value $\bra{0}\hat{H}(\tau_0)\ket{0}$, given a particular moment of time $\tau_0$, and subject to the normalization condition \eqref{normalization_cond}. Following the analysis done by Birrel and Davies \cite{BiDa84}, and with the aim of enabling a more direct comparison with renormalized approaches \cite{HaLaHo16}, we consider a similar procedure, the minimization of the expectation value of the energy density, that is $\bra{0}\hat{T}_{00}^{(f)}\ket{0}$.\\
	
	Analogously to the Hamiltonian diagonalization, the minimization of the energy density selects a particular vacuum state that minimizes the energy density expectation value associated to the fluctuations at a given time $\tau_0$. A clear advantage of this approach is that the expectation value $\bra{0}\hat{T}_{00}^{(f)}\ket{0}$ does not exhibit the divergence associated to the infinite volume of the space, which is present in the standard Hamiltonian diagonalization \cite{HaLaHo16, MuWi07, Fu79}. On the  other hand, as stated by Handley et al. this procedure still exhibits the usual ultraviolet divergence for large $k$ \cite{HaLaHo16}. Nonetheless, it is a good entry point in to more sophisticated approaches which consider renormalized energy-momentum tensors \cite{HaLaHo16,BDp, CoHo06}. \\
	
	Initially, in this approach, it is obtained the energy-momentum tensor of the fluctuations $f$ in the standard way
	\begin{equation}
		T_{\mu \nu}^{(f)} = \frac{-2}{\sqrt{-g}}\frac{\delta S_{(f)}}{\delta g^{\mu \nu}} = \frac{1}{a^2} \left\{f_{,\mu}f_{,\mu} - \frac{g_{\mu \nu}}{2} \left[ f_{,\alpha}f^{,\alpha} - \frac{z''}{z}f^2\right]\right\}.
	\end{equation}
	Therefore, the component $T_{00}$  takes the form
	\begin{equation}
		T_{00}^{(f)} = \frac{1}{2a^2}\left[\left(f'\right)^2 + \frac{1}{2} \left(\nabla f \right)^2 - \frac{z''}{z} f^2\right].
	\end{equation}
	Now, considering that the vacuum state satisfies $\hat{a}_k\ket{0} = 0,$ for all $\bar{k}$, and applying the expansion \eqref{fourier_f}, the expected density energy of the perturbation modes $f_k$ is written as
	\begin{equation}\label{exp_t_00}
		\bra{0}\hat{T}^{(f)}_{00}\ket{0} = \frac{1}{2a^2(\tau)} \int \frac{dk^3}{(2\pi)^{3}} \left[|f'_k|^2 + \left(k^2 - \frac{z''(\tau_0)}{z(\tau_0)}\right) |f_k|^2  \right].
	\end{equation}
	Now, we can  proceed to minimize the expectation \eqref{exp_t_00} with respect to the functions $f_k$ and $f'_k$, while obeying the normalization condition \eqref{normalization_cond}. Hence, we find that the solutions \eqref{sol_1} evaluated at a certain instant ($\tau_0$), must satisfy the conditions
	\begin{equation}\label{Q_conditions}
		\begin{aligned}
			|f_k|^2 &= \frac{1}{2 \sqrt{k^2 - \frac{z''}{z}}} \quad\text{and}\quad
			f'_k  &= - i \sqrt{k^2 - \frac{z''}{z}} f_k.
		\end{aligned}
	\end{equation}
	Specifically, for the De Sitter scale factor  \eqref{de_Sitter}, the conditions \eqref{Q_conditions} read
	\begin{equation}\label{dSitter_cond}
		\begin{aligned}
			|f_k|^2 &= \frac{1}{2 \sqrt{k^2 - \frac{2}{\tau^2}}} \quad\text{and}\quad
			f'_k  &= - i \sqrt{k^2 - \frac{2}{\tau^2}} f_k.
		\end{aligned}
	\end{equation}
	In the limit, $\tau \rightarrow -\infty$, corresponding to the distant past, the relations \eqref{dSitter_cond} become
	\begin{equation}\label{BD_cond}
		\begin{aligned}
			\lim_{\tau \to -\infty}	|f_k(\tau)|^2 &= \frac{1}{2k^2} \quad\text{and}\quad
			\lim_{\tau \to -\infty}	f'_k  &= - ik^2\,.
		\end{aligned}
	\end{equation}
	Additionally, applying the same limit to the solution \eqref{sol_1}, it is easy to see that the Bunch-Davies vacuum is recovered, that is, $A_k=1$ and $B_k = 0$. Therefore, for this vacuum choice, the solution is given by 
	\begin{equation}
		f_k(\tau) =  \frac{1}{\sqrt{2k}} \left(1 -\frac{i}{k \tau}\right) e^{-ik\tau}\,.
	\end{equation}
    Inserting $A_k=1$ and $B_k = 0$ in \eqref{ps_end} it reduces to
	\begin{equation}
		\left. P_{\delta \phi}(k) \right\rvert_{IR} = \frac{H^2}{4\pi^2}.
	\end{equation}
This power spectrum reflects the fact that the quantum inflaton fluctuations originate as nearly scale-invariant quantum vacuum fluctuations in a quasi-de Sitter background, where the Hubble parameter H is approximately constant during inflation.  This behavior is a generic prediction of single-field slow-roll inflationary models with canonical kinetic terms.
    
	\section{ Cut-off of quantum fluctuations during pre-inflation}

    Before exploring the behavior of this vacuum criterion in the inflationary case, we first analyze the dynamics of the quantum fluctuations $f_k$ in a simpler setting: a non-accelerating FLRW background. As we will show, the evolution of $f_k$ in this spacetime closely resembles the behavior of perturbations at the onset of inflation in the delayed-onset scenario ($t_0 \gg t_p$). This analogy provides useful intuition and simplifies the subsequent analysis of the de Sitter case.\\
	
	In the case of a non accelerated FLRW metric, and using conformal time, the scale factor has the form
	\begin{equation}\label{scale_factor_non_a}
		a(\tau) = a_0 \, exp\left(\frac{\tau}{\tau_0}\right).
	\end{equation} 
	
	Here $a_0$ and $\tau_0$ are the values of the scale factor and conformal time for an  arbitrary initial time $t_0$. It can be shown that for this background the Mukhanov equation \eqref{mukhanov_2} reads
	\begin{equation}
		f_{k}^{''} +\left(k^2 - \frac{1}{\tau_0^2}\right)f_k=0.
	\end{equation}
	This equation admits the following solutions
	\begin{equation}\label{sol_non_ac_I}
		f(\tau) = A_k e^{-i\omega(k)\tau} + B_k e^{i\omega(k)\tau} \quad \text{for} \quad k>q_{0}^2,
	\end{equation}
	\begin{equation}\label{sol_non_ac_II}
		f(\tau) = A_k e^{\Omega(k)\tau} + B_k e^{\Omega(k)\tau} \quad \text{for} \quad k<q_{0}^2\,,
	\end{equation}
	where $\omega(k) = \sqrt{k^2-q_0^2}$, $\Omega(k) = \sqrt{q_0^2- k^2}$ and $q_0 = 1/\tau_0$. In order to determine the free constants $A_k$ and $B_k$ in \eqref{sol_non_ac_I} and \eqref{sol_non_ac_II}, we will  apply the energy density minimization conditions \eqref{Q_conditions}. For this scenario, the normalization conditions \eqref{Q_conditions} become
	\begin{equation}\label{non_a_tau_cond}
		\begin{aligned}
			|f_k|^2 &= \frac{1}{2 \sqrt{k^2 - \frac{1}{\tau_0^2}}} \quad\text{and}\quad
			f'_k  &= - i \sqrt{k^2 - \frac{1}{\tau_0^2}} f_k.
		\end{aligned}
	\end{equation}
	Substituting the normalization relations \eqref{non_a_tau_cond} into the energy density expectation value \eqref{exp_t_00} we obtain
	\begin{equation}\label{exp_t_00_II}
		\bra{0}\hat{T}^{(f)}_{00}(\tau_0)\ket{0} = \frac{1}{2a^2} \int \frac{dk^3}{(2\pi)^{3}}  \,k\sqrt{1-\frac{q_0^2}{k^2}}.
	\end{equation}
	
	From \eqref{exp_t_00_II}, it is evident that the energy density becomes complex for values $k<q_0$. In consequence, the normalization expressions \eqref{non_a_tau_cond} are valid only for modes with $k>q_0$. Hence, for modes obeying  $k<q_0$ this vacuum criteria cannot be applied, suggesting that the vacuum state associated with these perturbations differs from that of the modes with $k>q_0$. 
	Applying the normalization conditions \eqref{non_a_tau_cond} to the solutions \eqref{sol_non_ac_I}, we find that the constants $A_k$ and $B_k$ for this scenario take the form 
	\begin{equation}\label{ctes_non_acc_modes}
		\begin{aligned}
			A_k = \frac{1}{\sqrt{2\left(k^2 - q_0^2\right)}}  \quad\text{and}\quad
			B_k  &=  0.
		\end{aligned}
	\end{equation}
	
	Furthermore, in the non accelerated scenario, the $z''/z$ term  is related to the comoving Hubble parameter ($\mathcal{H} = a'/a$) as follows
	\begin{equation}
		\frac{z''}{z}=\mathcal{H}^2 = \frac{1}{\tau^2_0}.
	\end{equation} 
	Therefore, we can conclude that for modes with an associated wavelength greater than the Hubble radius, this vacuum selection criteria breaks down, and thus,  only modes inside the Hubble sphere can admit the normalization conditions \eqref{non_a_tau_cond}. Moreover, for modes well inside in the Hubble sphere ($k^2 \gg \mathcal{H}$), the constants \eqref{ctes_non_acc_modes} tend towards the "Bunch-Davies" values ($A_k=1$ $B_k=0$).  As will be shown in the next section, an analogous situation arises in a  delayed-onset scenario of inflation.

	\section{Presence of a cutt off in the power-spectrum due to a delayed onset of Inflation}
	Now that the formalism has been established, we examine its implications for a delayed onset of inflation. A noteworthy implication of the vacuum criteria reviewed in the previous sections is that, for the case where inflation starts at a time $t_0 > t_p$, we find that this approach breaks down for modes $f_k$ with $k<2/\tau^2(t_0)$ in a similar manner that happens in a non accelerated scenario. Identifying the vacuum state as the one that minimizes the energy density expectation value \eqref{exp_t_00}, at the start of inflation $(\tau = \tau_0)$ and considering the Mukhanov equation \eqref{mukhanov_3}, we find that the normalization conditions \eqref{dSitter_cond} are accordingly modified and take the form
	\begin{equation}\label{tau_0_cond}
		\begin{aligned}
			|f_k|^2 &= \frac{1}{2 \sqrt{k^2 - \frac{2}{\tau_0^2}}} \quad\text{and}\quad
			f'_k  &= - i \sqrt{k^2 - \frac{2}{\tau_0^2}} f_k.
		\end{aligned}
	\end{equation}
	Imposing these conditions on the general solution \eqref{sol_1} for the $f_k$ modes in a De Sitter epoch, it can be shown that the constants $A_k$ and $B_k$ are expressed in this case as 
	\begin{equation*}\label{norm_a_b_sitter}
		A_k = \frac{k}{2} \sqrt{\frac{\left(1+\frac{\omega(k)}{k}\right)^2 + \frac{q_0^2}{2k^2}}{\omega(k)}},
	\end{equation*}
	\begin{equation}
		B_k  = - \frac{k}{2} \sqrt{\frac{\left(1+\frac{\omega(k)}{k}\right)^2 + \frac{q_0^2}{2k^2}}{\omega(k)}}\left[ \frac{i\omega(k)\left(1-\frac{iq_0}{\sqrt{2}k}\right)-\frac{q_0}{\sqrt{2}}\left(1-\frac{iq_0}{\sqrt{2}k}\right)-ik}{i\omega(k)\left(1+\frac{iq_0}{\sqrt{2}k}\right)-\frac{q_0}{\sqrt{2}}\left(1+\frac{iq_0}{\sqrt{2}k}\right)+ik}\right],
	\end{equation}
	where $q_0^2 = 2/\tau_0^2$ and $\omega(k) = \sqrt{k^2 - q_0^2}$. On the other hand, substituting the relations \eqref{tau_0_cond} in the energy density expectation value, we obtain the following
	\begin{equation}\label{exp_t_00_III}
		\bra{0}\hat{T}^{(f)}_{00}(\tau_0)\ket{0} = \frac{1}{2a^2} \int \frac{dk^3}{(2\pi)^{3}}  \,k\sqrt{1-\frac{q_0^2}{k^2}}.
	\end{equation}
	Additionally, for this scenario we have that the term $z''/z$ in the Mukhanov equation \eqref{mukhanov_3} relates to the comoving Hubble parameter as 
	\begin{equation}\label{factor_z_sitter}
		\frac{z''}{z}= \frac{2}{\tau^2} \approx 2 \mathcal{H}^2.
	\end{equation}
	The results \eqref{exp_t_00_III} and \eqref{factor_z_sitter} resemble the behavior seen in the non accelerated scenario. The difference is that for the De Sitter case, the comoving Hubble parameter evolves, while in the non accelerated case, this quantity remains constant. Hence, at the beginning of inflation, conclusions analogous to those of the non accelerated expansion can be drawn. Consequently, for modes $k < q_0$, the energy density expectation value in $t_0$ becomes complex, and therefore this vacuum selection criterion breaks down under these conditions. Taking into account the previous analysis and the power spectrum \eqref{ps_end}, one can conclude that this criterion can explain the possible existence of a cut-off in the power spectrum of inflaton fluctuations.\\

	Furthermore, in the limit $k \gg q_0$, that is, when the modes are well inside the Hubble sphere,  the normalization constants \eqref{norm_a_b_sitter} take the values $A_k \approx 1$ and $B_k \approx 0$. Therefore, the vacuum states of the modes $f_k$ with $k \gg q_0$ are  well approximated by a Bunch-Davies vacuum in this situation. 
	
	\section{\large{Unified Pre-inflationary-inflationary Model}}
	
    Now we are in position to construct a unified model of a pre-inflationary epoch followed by an inflationary era.  In both regimes, the dynamics are assumed to be dominated exclusively by a vacuum energy component modeled by a scalar field. During the pre-inflationary epoch, this field will be referred to as the pre-inflaton, while in the inflationary epoch it will be identified as the inflaton.\\
	
	On the other hand, it is well-known in the literature, when we speak about cosmic inflation,   that there are basically two ways of constructing an inflationary model: namely by choosing a potential for the inflaton field or by selecting a scale factor. In this paper we will adopt the second criterion i.e. we first implement a phenomenological choosing of the scale factor and then we will determine the potential for the pre-inflaton  and inflaton  fields according to the field equations of the model.\\

    Before introducing a phenomenological scale factor, it is instructive to recall a standard assumption in conventional inflationary scenarios: all quantum modes associated with the inflaton field are initially well within the Hubble horizon at the onset of inflation. Consequently, at early times these modes are causally connected and thus spatially correlated. As inflation proceeds, the comoving Hubble radius decreases, leading modes to cross outside the horizon, after which their amplitudes effectively freeze and undergo decoherence.

From this viewpoint, if one extrapolates this picture to a pre-inflationary stage characterized by a similarly decreasing horizon, it would suggest that the modes have remained correlated since very early times. However, this expectation may be in tension with the observed suppression of correlations at large angular scales (above $\sim 60^\circ$) in the cosmic microwave background, as indicated by \textit{Planck} observations \cite{PlanckVII}.\\

Hence, a possible way to alleviate this tension is to consider a pre-inflationary stage in which the Hubble horizon increases with time. Such a behavior naturally arises if the expansion of the Universe is decelerating during this epoch, thereby allowing previously disconnected regions to come into causal contact. Thus, in order to concentrate all these ideas we consider  a power law cosmic scale factor of the form
	\begin{equation}\label{factor_1}
		a(t) = a_{pl}\left[ \frac{1+(1-\beta)\alpha t}{1+(1-\beta)\alpha t_{pl}}\right]^{\frac{1}{1-\beta}},
	\end{equation}
	where $t_{pl}$ is denoting the Planckian time,  $a_{pl} = a(t_{pl})$, and $\alpha$, $\beta$ are parameters of the model which can take different values during the pre-inflationary and in\-fla\-tio\-na\-ry epochs. Restrictions on these parameters will be determined by the model. The corresponding Hubble parameter coming from \eqref{factor_1} is then 
	\begin{equation}\label{Hubble_1}
		H(t) = \frac{\alpha}{1+\alpha (1-\beta)t}.
	\end{equation}  
    In order to construct a consistent pre-inflationary model, it should provide a reliable approximation to the Planck era. At Planck time, defined as $t_{\mathrm{pl}} = \ell_{\mathrm{pl}}/c$, the following relations are expected to hold
	\begin{equation}\label{prel}
		l_{pl}=\frac{\hbar}{c m_p},\qquad G=\frac{c\hbar}{2m_p^2},
	\end{equation}
	where $m_p \approx 1.2209\times 10^{19}\,GeV$   and $l_p \approx 1.616\times 10^{-35}\,m\approx 8.1\times 10^{-20}\,GeV^{-1}$ are the Planck mass and length respectively.\\
	
	The energy density in this stage calculated at Planck's time should be equal to the Planck's energy density given by
	\begin{equation}\label{planck_rho}
		\rho_{pl} = \frac{m_p}{4/3 \pi l_p^3}=\frac{3}{8\pi G}\frac{1}{t_{pl}^2}.
	\end{equation}
	On the other hand, considering \eqref{friedmann_1} the Hubble parameter \eqref{Hubble_1} leads to an energy density
	\begin{equation}\label{inipla1}
		\rho(t)=\frac{3}{8\pi G} \frac{\alpha^2}{[1+\alpha (1-\beta)t]^2}.
	\end{equation}
	It is not difficult to verify from \eqref{inipla1}  that $\rho(t_{pl})$ matches the Planck energy density $\rho_{pl}$ when $\alpha=(\beta t_{pl})^{-1}$. Therefore, adopting a scale factor of the form \eqref{factor_1} during the pre-inflationary epoch provides a natural mechanism by which the energy density evolves smoothly toward the Planck scale, ensuring a consistent connection with the expected physics of the Planck era.\\

Now, let us consider a semiclassical approximation for the pre-inflaton field, ana\-lo\-gous to the form given in Eq.~(10), but expressed in terms of the cosmic time $t$. In this framework, the Friedmann equation (5) implies that the Hubble parameter can be decomposed as $H(x^\sigma) = H_b(t) + \delta H(x^\sigma)$, where $H_b(t)$ represents the background Hubble parameter, valid on cosmological (large) scales, and $\delta H(x^\sigma)$ encodes the local fluctuations induced by the quantum fluctuations of the pre-inflaton field. It then follows from Eqs.~(5), (6), (7), and (10) that, on large scales, the dynamical equations take the form
	\begin{eqnarray}\label{ced1}
		&& H_b^2=\frac{M_{pl}^2}{3}\left(\frac{1}{2}\dot{\phi}_b^2+U(\phi_b)\right),\\
		\label{ced2}
		&& \ddot{\phi}_b+3H_b\dot{\phi}_b+\left.\frac{dU}{d\phi}\right|_{\phi_b}=0,
	\end{eqnarray}
	where $M_{pl}=(8\pi G)^{-1/2}$ is the reduced planckian mass. Now, by deriving \eqref{ced1} with respect the cosmic time $t$ and employing \eqref{ced2} we arrive to
	\begin{equation}\label{ced3}
		\dot{\phi}_b^2=-2M_{pl}^2\dot{H}_b\,.
	\end{equation}
	Inserting \eqref{ced3} in \eqref{ced1} we obtain 
	\begin{equation}\label{ced4}
		U_b=M_{pl}^2\left(3H_b^2+\dot{H}_b\right).
	\end{equation}
	With the help of \eqref{Hubble_1} and \eqref{ced4}, it is not difficult to show that the potential corresponding to the scale factor \eqref{factor_1} reads
	\begin{equation}\label{U_phi}
		U_b(\phi_b) = \frac{\alpha^2 M^2_{pl} (\beta+2)}{(1+\alpha(1-\beta)t_p)^2} \exp\left( \frac{-\sqrt{2(1-\beta) }\,\phi_b}{M_{pl}} \right).
	\end{equation}
	It can be verified by simple inspection that for having $U_b\geq 0$, it is required that $-2 \leq \beta \leq 1$. For $\beta=-2$ the potential becomes zero and thus the expansion will be kinetically dominated. For $\beta=1$ the potential becomes constant and thus the expansion will be De-Sitter.\\

	With the aim of analyzing the dynamical behavior of the expansion, we will examine the behavior of the comoving Hubble radius $\mathcal{R}_H = 1/aH$. For the   scale factor \eqref{factor_1} it reads
	\begin{equation}\label{hubble_radius}
		\mathcal{R}_H = \frac{\left( 1+ \alpha (1-\beta) t_p\right)}{\alpha a_p} \left(\frac{1+(1-\beta)\alpha t}{1+(1-\beta)\alpha t_p}\right)^{\frac{\beta}{\beta-1}}.
	\end{equation}
	In this manner, based on the Hubble radius \eqref{hubble_radius} and the fact that the  deceleration parameter is  $q = -\beta$ for \eqref{factor_1} we can identify four different dynamical stages depending of the allowed values of $\beta$. These cases can be categorized as follows:

	\begin{itemize}
		\item Phase I $(-2 \leq \beta < 0)$: For each one of these values of $\beta$ the expansion is decelerated, and the comoving Hubble radius  $\mathcal{R}_H$ increases in size.
		\item Phase II $(\beta = 0)$: In this, case the expansion does not accelerate $(q = 0)$, and $\mathcal{R}_H$ remains constant.
		\item Phase III $(0 < \beta < 1)$: Here the expansion is accelerated and    $\mathcal{R}_H$ decreases in size.
		\item Phase IV $(\beta = 1)$. At this stage, the expansion corresponds to a De Sitter, during which $\mathcal{R}_H$ remains constant.
	\end{itemize}
	
	Thus, when the parameter $\beta$ shifts from values in the interval $(-2 \leq \beta < 0)$
	to values in $0\leq\beta<1$, the universe transitions from a phase of decelerated expansion to a phase of accelerated expansion, ultimately approaching a De Sitter regime. The transition point between deceleration and acceleration occurs at $\beta=0$. Therefore, the decelerated phase I $(-2 \leq \beta < 0)$ can be associated with a preinflationary epoch, while the accelerated Phase III $(0 < \beta < 1)$ will correspond to the inflationary period of the Universe. The Phase II $(\beta = 0)$ represents a transition era connecting the pre-inflationary and inflationary stages.  During the transition from decelerated to accelerated expansion, the exponent of the scale factor in  \eqref{factor_1} is $p=1/(1-\beta)=1$.
	Hence, during the inflationary period, the exponent satisfies $p\in(1,\infty)$. In the limit $p\to\infty$, the expansion approaches a de Sitter regime. We refer to this initial phase of inflation as soft inflation. A second inflationary stage follows, beginning with de Sitter-like expansion and ending in a power-law expansion. This later phase can be associated with the inflationary epoch described by standard models.
	\\

	One important quantity in inflationary models is the number of e-folds $N$. During inflation, $N$ provide a measure of the exponential expansion of the Universe. Physically, one e-fold corresponds to an increase of the scale factor by a factor of $e$, so that the total number of e-foldings quantifies how much the Universe grows during inflation. This quantity is typically defined as $N \equiv \ln\!\left(a/a_i\right)$, where $a_i$ is the initial value of the scale factor. A sufficiently large number of e-foldings (usually $N \gtrsim 50\text{--}60$) ensures that inflation can successfully address the horizon and flatness problems by stretching initially microscopic, causally connected regions to cosmological scales. Mathematically $N$ is defined as 
	\begin{equation}
		N \equiv \int_{a_0}^{a_f} d\left(ln(a)\right)= \int_{\phi_0}^{\phi_f} \frac{H}{\dot{\phi}} d\phi =  \int_{t_0}^{t_f} H dt,
	\end{equation}
	Employing \eqref{Hubble_1} and \eqref{ced3}  the number of e-foldings in our particular model as a function of field values at the beginning and end of the inflationary period results to be 
	\begin{equation}
		N = \sqrt{4\pi G} \left(\frac{\phi_f}{1-\beta_f} - \frac{\phi_0}{1-\beta_0}\right).
	\end{equation}
	and as a function of the time values at the beginning and end of inflation we obtain
	\begin{equation}
		N = \frac{1}{1-\beta_f}\ln \left(1+(1-\beta_f)\alpha_f t_f\right) - \frac{1}{1-\beta_0}\ln \left(1+(1-\beta_0)\alpha_0 t_0\right),
	\end{equation}
	Furthermore, it follows from \eqref{ced3} that  the behavior of  $\phi_b$ for phases with $\beta \neq 0$, is given by
	\begin{equation}
		\phi_b(t) = \phi(t_0) + M_p\sqrt{\frac{2}{1-\beta}} \ln\left(\frac{1+\alpha(1-\beta)t}{1+\alpha(1-\beta)t_p}\right),
	\end{equation} 
	where $t_0$ is the initial time of the studied phase. For the transition phase with $\beta = 0$, the field $\phi_b$ acquires the form
	\begin{equation}
		\phi_b(t) = \phi(t_0) + M_p\sqrt{2} \ln\left(\frac{1+\alpha t}{1+\alpha t_p}\right).
	\end{equation}
    The former two expressions determine the dynamics of the  pre-inflaton and the inflaton fields at large cosmological scales. However, the dynamical behavior of the quantum fluctuations associated with both the pre-inflaton and inflaton fields remains to be determined. This issue will be addressed in detail in the following section.

	\section{\large{Dynamics of the quantum fluctuations of a scalar field during pre-inflation and inflation}}

    In order to analyze the scalar fluctuations of the inflaton field generated during both the pre-inflationary and inflationary stages within this unified framework, we will employ the Fourier expansion \eqref{fourier_f} together with the Mukhanov equation \eqref{mukhanov_2}. Since the Mukhanov equation is formulated in terms of conformal time $\tau$, it is necessary to first determine its explicit form across the different phases of the cosmic evolution, from pre-inflation to inflation, within our model. \\

    With this idea in mind,  substituting \eqref{factor_1} in \eqref{conformal_time} for $\beta \neq 0$, we obtain 
	\begin{equation}\label{ctime1}
		\tau = \frac{-\left(1+(1-\beta)t_p\right)}{\alpha \beta a_p} \left(\frac{1+(1-\beta)\alpha t}{1+(1-\beta)\alpha t_p}\right)^{\frac{\beta}{\beta-1}}.
	\end{equation}
	In contrast, when $\beta = 0$, the same calculation gives
	\begin{equation}\label{ctime2}
		\tau = \frac{1+\alpha t_p}{\alpha a_p}\ln\left(1+\alpha t\right).
	\end{equation}
Hence, it can be verified, that in view of the Mukhanov equation \eqref{mukhanov_2} and the expressions \eqref{ctime1}
 and \eqref{ctime2},   the dynamics of the modes $f_k$ for our  pre-inflationary-inflationary model is dictated by the equation 
	\begin{equation}\label{mukhanov_4}
		f_k^{''} +\left(k^2  - \frac{\beta +1}{\beta^2 \tau^2}\right)f_k  = 0.
	\end{equation}
    This equation has for solution
	\begin{equation}\label{f_k_i}
		f_k (\tau) = \sqrt{-\tau} \left(C_1 \mathrm{H}_{\nu}^{(1)}(-k\tau) + C_2 \mathrm{H}_{\nu}^{(2)}(-k\tau)\right),
	\end{equation}
	where $\mathrm{H}_{\nu}^{(1)}(y)$ and $\mathrm{H}_{\nu}^{(2)}(y)$ are the Hankel functions of first and second species, and $\nu = (2+\beta)/2\beta$.
	\\	
	
	On the other hand, expressing the scale factor \eqref{factor_1} in terms of the conformal time \eqref{ctime1}, the comoving Hubble parameter for $\beta \neq 0 $ reads
	\begin{equation}\label{cHubble}
		\mathcal{H} = - \frac{1}{\beta \tau}.
	\end{equation}
	Notice that in the limit $\beta \to 0$, the comoving Hubble factor tends to the number
	\begin{equation}\label{chubble0}
		\mathcal{H}^2_0 = \frac{\alpha^2 a_p^2}{(1+\alpha t_p)^2}.
	\end{equation}
By introducing the comoving Hubble parameter \eqref{cHubble}, Eq.~\eqref{mukhanov_4} can be recast in the form
\begin{equation}\label{qcya}
    f_k'' + \left[k^2 - (\beta + 1)\mathcal{H}^2(\tau)\right] f_k = 0.
\end{equation}
The general behavior of the solutions of \eqref{qcya} depends on the evolution of $\mathcal{H}^2(\tau)$ and, for a given wavenumber $k$, allows us to distinguish the following processes: 
\begin{itemize}
    \item \textbf{ A Decoherence Process:} If $\mathcal{H}$ increases with conformal time $\tau$, then at early times one has $k \gg \mathcal{H}$, so that the $k^2$ term dominates and the mode equation reduces to
    \begin{equation}
        f_k'' + k^2 f_k = 0.
    \end{equation}
    As the Universe evolves, $\mathcal{H}$ grows and eventually the condition $k \ll \mathcal{H}$ is reached. In this regime, the effective mass term dominates, leading to
    \begin{equation}
        f_k'' - (\beta + 1)\mathcal{H}^2(\tau) f_k = 0.
    \end{equation}
Thus, in this case, the modes will pass form being coherent to the decoherence. 
    \item \textbf{A Coherence Process:} If $\mathcal{H}$ decreases with $\tau$, then for a given mode one initially has $k^2 \ll \mathcal{H}^2$, and the dynamics is governed by
    \begin{equation}
        f_k'' - (\beta + 1)\mathcal{H}^2(\tau) f_k = 0.
    \end{equation}
    As time evolves and $\mathcal{H}$ decreases, the condition $k^2 \gg \mathcal{H}^2$ is eventually satisfied, and the equation becomes
    \begin{equation}
        f_k'' + k^2 f_k = 0.
    \end{equation}
Therefore, in this case, the modes evolve from a decoherent regime to a coherent one
    \item \textbf{A Static Process:} If $\mathcal{H}$ remains constant, the behavior of the modes is fixed by their wavenumber. Modes with $k^2 > \mathcal{H}_0^2$ remain oscillatory (coherent), whereas modes with $k^2 < \mathcal{H}_0^2$ are dominated by the effective mass term and exhibit decoherence.
\end{itemize}
In our pre-inflationary-inflationary model described above, the different dynamical phases are characterized by distinct behaviors of $\mathcal{H}^2(\tau)$. For the phase I ($-1 \leq \beta < 0$), we obtain
\begin{equation}
    \mathcal{H}^2(\tau(t)) = \frac{\alpha^2 a_p^2}{\left(1+\alpha(1-\beta)t_p\right)^2}
    \left(\frac{1+\alpha(1+|\beta|)t_p}{1+\alpha(1+|\beta|)t}\right)^{\frac{2|\beta|}{|\beta +1|}}.
\end{equation}
From this expression, it follows that $\mathcal{H}^2$ decreases with cosmic time $t$ during phase I, indicating the presence of a coherence process. Consequently, inflaton fluctuations do not freeze out during the pre-inflationary stage; instead, as the universe expands, these modes effectively re-enter the horizon and exhibit an oscillatory behavior (a coherence process).\\

During the phase II ($\beta = 0$), $\mathcal{H}_0^2$ is given by \eqref{chubble0} and remains constant. In this case, neither coherence nor decoherence processes take place. Modes with $k^2 \gg \mathcal{H}_0^2$ remain oscillatory, whereas modes with $k^2 \ll \mathcal{H}_0^2$ are dominated by the effective mass term and become frozen.\\

Meanwhile, during the phase III, i.e,  for $0 < \beta < 1$, $\mathcal{H}$ is given by
\begin{equation}
    \mathcal{H}^2(\tau(t)) = \frac{\alpha^2 a_p^2}{\left(1+\alpha (1-\beta)t_p\right)^2}
    \left(\frac{1+\alpha (1-\beta)t}{1+\alpha (1-\beta)t_p}\right)^{\frac{2\beta}{1-\beta}}.
\end{equation}
In this regime, $\mathcal{H}^2$ grows with time, signaling the onset of a decoherence process. As a result, inflaton quantum perturbations are driven outside the horizon and freeze out, just as happens in standard inflationary scenarios.\\

Now, in order to complete the analysis of the quantum perturbations of the inflaton field, it is necessary to implement the canonical quantization procedure. This allows us to identify the appropriate vacuum state and, consequently, to fix the integration constants appearing in Eq.~\eqref{f_k_i}.
Adopting the procedure introduced on Section 2.1 and 2.2, we
obtain for our model that the expected energy density of the perturbation modes $f_k$ results to be
	\begin{equation}\label{exp_t_00_2}
		\langle 0 | \hat{T}^{(f)}_{00} (\tau_0) | 0 \rangle  = \frac{1}{2a^2(\tau_0)} \int \frac{dk^3}{(2\pi)^{3}} \left[|f'_k(\tau_0)|^2 + \left(k^2 - \frac{(\beta_0+1)}{\beta_0^2\tau_0^2}\right) |f_k(\tau_0)|^2  \right].
	\end{equation}
Following the diagonalization procedure of $\hat{T}^{(f)}_{00}$, we minimize the expectation value \eqref{exp_t_00_2}, treating it as a functional of $f_k$, $f_k^{*}$, $f_k'$, and $f_k^{\prime *}$ at the onset of inflation ($\tau_0$), subject to the normalization condition \eqref{normalization_cond}. This procedure leads to the following conditions for the mode functions \eqref{f_k_i} at $\tau_0$
\begin{equation}
    |f_k(\tau_0)|^2 = \frac{1}{2 \sqrt{k^2 - \frac{\beta_0+1}{\beta_0^2 \tau_0^2}}},
\end{equation}
\begin{equation}\label{normalization_condII}
    f_k'(\tau_0) = - i \sqrt{k^2 - \frac{\beta_0+1}{\beta_0^2 \tau_0^2}} \, f_k(\tau_0).
\end{equation}

It is important to note that for modes satisfying $k^2 < (\beta_0+1)\mathcal{H}^2$, the expectation value of the energy density $\langle 0 | \hat{T}^{(f)}_{00} (\tau_0) | 0 \rangle$ becomes complex. Therefore, the nor\-ma\-li\-za\-tion condition \eqref{normalization_condII} is only well-defined for modes with $k^2 > (\beta_0+1)\mathcal{H}^2$. This indicates that the vacuum prescription adopted here applies exclusively to sub-horizon modes, while super-horizon modes require a different treatment, suggesting a distinct vacuum structure for these perturbations.

Imposing these conditions on the solutions \eqref{f_k_i} at $\tau_0$, for a general $\beta_0$, we obtain the normalization constants $C_1$ and $C_2$ as
\begin{equation*}
    C_1 = \left\{ -2 \sqrt{(k\tau_0)^2 - \frac{\beta_0 +1}{\beta_0^2}} \left[\left(1 + \frac{|\Delta|^2}{|\delta|^2}\right)\mathrm{H}_{\nu}^{(1)}\mathrm{H}_{\nu}^{(2)} - \frac{\Delta^*}{\delta^*} \left(\mathrm{H}_{\nu}^{(1)}\right)^2 - \frac{\Delta}{\delta} \left(\mathrm{H}_{\nu}^{(2)}\right)^2\right]\right\}^{-1/2},
\end{equation*}
\begin{equation}\label{c_1_c_2}
    C_2 = -\frac{\Delta}{\delta} \, C_1,
\end{equation}
where $\mathrm{H}_{\nu}^{(1)} = \mathrm{H}_{\nu}^{(1)}(-k\tau_0)$, $\mathrm{H}_{\nu}^{(2)} = \mathrm{H}_{\nu}^{(2)}(-k\tau_0)$, $\nu = (2 + \beta_0)/(2\beta_0)$, and the coefficients $\Delta$ and $\delta$ are given by
\begin{equation*}
    \Delta(\beta_0) = \sqrt{- \tau_0}\left[\frac{1}{\tau_0} \left( \frac{1}{2} - \nu\right) + i \sqrt{k^2 - \frac{\beta_0 + 1}{\beta_0^2 \tau_0^2}}\right]\mathrm{H}_{\nu}^{(1)}(-k\tau_0) - k \sqrt{-\tau_0} \, \mathrm{H}_{\nu-1}^{(1)}(-k\tau_0),
\end{equation*}
\begin{equation}
    \delta(\beta_0) = \sqrt{- \tau_0}\left[\frac{1}{\tau_0} \left( \frac{1}{2} - \nu \right) + i \sqrt{k^2 - \frac{\beta_0 + 1}{\beta_0^2 \tau_0^2}}\right]\mathrm{H}_{\nu}^{(2)}(-k\tau_0) - k \sqrt{-\tau_0} \, \mathrm{H}_{\nu-1}^{(2)}(-k\tau_0).
\end{equation}

It is worth noting that n the asymptotic limit $-k\tau_0 \to \infty$, these expressions reduce to the standard Bunch--Davies vacuum, namely $C_1 = \sqrt{\pi}/2$ and $C_2 = 0$. Additionally, we remark that these results are valid only for modes satisfying $k^2 > (\beta_0+1)\mathcal{H}^2$ at the onset of inflation.\\

Taking into account the behavior of the comoving Hubble horizon discussed above, and using the constants \eqref{c_1_c_2} together with the expression for the power spectrum at the end of inflation \eqref{power_II}, we obtain that the infrared contribution to the power spectrum of the inflaton fluctuations $\delta\phi$, for a general initial phase characterized by $\beta_0$ and a final accelerated phase with $0 \leq \beta < 1$, is given by
\begin{equation}
    \left. P_{\delta \phi}(k,\beta, \beta_0) \right\rvert_{IR} =
    \frac{2^{2\nu}\left|C_1(\beta_0)\right|^2 k^{3-2\nu}\Gamma^2(\nu)}{(2\pi^5)a_p^2}
    \left(\frac{\alpha\beta a_p}{1+\alpha(1-\beta)t_p}\right)^{2/\beta}
    \left(1 + \frac{\Delta\delta^{*}+\Delta^{*}\delta}{|\delta|^2} + \frac{|\Delta|^2}{|\delta|^2}\right),
\end{equation}
where $\nu(\beta) = (2+\beta)/(2\beta)$. A nearly scale-invariant power spectrum corresponds to $\nu \approx 3/2$, which in turn implies $\beta \approx 1$. Observational constraints from the 2018 \textit{Planck} data release indicate that the scalar spectral index lies in the range $n_s = 0.9626 \pm 0.0057$ \cite{planck_vi}. Using the relation $n_s - 1 = 3 - 2\nu$, this translates into $\beta = 0.9816 \pm 0.0028$ during the inflationary epoch.

Furthermore, in the limit $\beta \approx 1$ and $\beta_0 \to 0$, one finds that $C_1(0) = \sqrt{\pi}/4$ and $\Delta = 0$. In this case, corresponding to an initial non-accelerated phase, the power spectrum simplifies to
\begin{equation}
    \left. P_{\delta \phi}(k,\beta \approx 1, \beta_0 = 0) \right\rvert_{IR} = \frac{\alpha}{4\pi^2}.
\end{equation}
This result coincides with the standard expression obtained for an inflationary phase described by a de Sitter expansion.

\section{\large{Final discussion}}

In this paper, we have proposed a novel unified cosmological model that consistently describes a pre-inflationary epoch followed by an inflationary phase, both driven by a single scalar field. Unlike conventional approaches that largely ignore the pre-inflationary dynamics, in our model, the early Universe begins at the Planck time with a decelerated expansion, characterized by an increasing comoving Hubble horizon, and subsequently undergoes a smooth transition toward an accelerated inflationary regime.\\

A key feature of the model is that the pre-inflationary phase allows for a causal structure different from that of standard inflation. In particular, the growth of the Hubble horizon during this stage permits the existence of modes that remain causally disconnected at early times. This provides a natural mechanism to account for the observed suppression of correlations at large angular scales $(\theta \gtrsim 60^\circ)$ in the CMB, as reported by \textit{Planck} data.\\

We have shown that the adoption of a phenomenological scale factor enables a smooth interpolation between the Planck era and the inflationary phase, ensuring consistency with the expected Planck-scale energy density. Furthermore, by implementing a semiclassical treatment of the scalar field dynamics, we derived the corresponding effective potential and established the conditions under which the model reproduces the desired cosmological behavior.\\

The analysis of quantum fluctuations was carried out using the Mukhanov--Sasaki formalism together with a vacuum selection criterion based on the minimization of the energy density expectation value. This approach allowed us to identify a natural limitation of the vacuum prescription: it is only well-defined for sub-horizon modes. As a consequence, super-horizon modes require a different treatment, leading to a modified structure of the initial quantum state. An important outcome of this analysis is the emergence of a cutoff in the primordial power spectrum. This cutoff arises from the breakdown of the vacuum selection criterion for modes with $k^2 < (\beta_0 + 1)\mathcal{H}^2$ at the onset of inflation. We showed that this feature is consistent with scenarios involving a delayed onset of inflation and provides a potential explanation for the lack of angular correlations at large scales. In other words, at the perturbative level, the implementation of canonical quantization together with an energy-minimization vacuum prescription reveals a crucial and novel feature: the vacuum state is only well-defined for sub-horizon modes at the onset of inflation. This naturally induces a cutoff in the primordial power spectrum, without the need to introduce it by hand. Such a cutoff emerges as a direct consequence of the underlying physics and leads to a modified spectrum at large scales.

Finally, we derived the general expression for the power spectrum of inflaton fluctuations in our model and showed that, in the appropriate limits, it reduces to the standard nearly scale-invariant result. In particular, for $\beta \approx 1$ and $\beta_0 \to 0$, the spectrum reproduces the well-known de Sitter result, ensuring consistency with observational constraints.\\

Overall, the model presented here provides a coherent extension of inflationary cosmology toward earlier times to address some of the open issues in standard inflationary cosmology, particularly those related to initial conditions and large-scale anomalies in the CMB. Further work could explore more refined vacuum prescriptions as well as possible observational signatures that distinguish this scenario from conventional inflationary models.\\

	\section*{\large{Acknowledgements}}
	
	\noindent J. E. Madriz-Aguilar, A. Bernal and M. Montes acknowledge  SECIHTI
	(M\'exico) and Centro Universitario de Ciencias Exactas e Ingenierias of Guadalajara University for financial support. Diego Allan Reyna   thanks SECIHTI
	(M\'exico) for
financial support. 
	\bigskip
	
	\section*{\large{Data Availability Statement}}
	The data that support the findings of this study are available at 
	https://doi.org/10.1051/0004-6361/201833910.

\end{document}